\documentclass[%
 reprint,
 superscriptaddress,
 amsmath,amssymb,
 aps,
]{revtex4-1}

\usepackage{graphicx}
\usepackage{dcolumn}
\usepackage{bm}

\usepackage{upgreek}

\begin{document}

\preprint{APS/123-QED}

\title{Imaging through Fano-resonant dielectric metasurface governed by quasi-BIC}

\author{Chaobiao Zhou}
\email{cbzhou@gzmu.edu.cn}
\affiliation{College of Mechanical and Electronic Engineering, Guizhou Minzu University, Guiyang 550025, China}

\author{Shiyu Li}
\affiliation{Wuhan National Laboratory for Optoelectronics, Huazhong University of Science and Technology, Wuhan 430074, China}


\begin{abstract}
Fano resonance has raised great attention in nanophotonics attributing to its unique properties. In this work, we study the imaging function of Fano-resonant silicon metasurfece governed by quasi-bound states in the continuum (BICs). First, by breaking the in-plane symmetry of nanodisks, a symmetry-protected quasi-BIC is excited with the emergence of a sharp Fano resonance. The near field distributions, multipole contributions and radiation patterns of the metasurface are performed to uncover the mechanism and characteristics of this resonance. In addition, we investigate the imaging function of this Fano-resonant metasurface assisted by phase-change material $\rm Ge_2Sb_2Te_5$ (GST). Through selective modification of different units from $a$-GST to $c$-GST, the produced transmitted image well reconstructs the target letter. Our finding may provide a route to achieve efficient metasurface-based imaging and fast spatial modulations.
\end{abstract}

\maketitle


\section{\label{sec:1}Introduction}
Resonant metasurfaces have received extensive attention due to their sharp spectral feature and extraordinary field enhancement \cite{miroshnichenko2010fano,limonov2017fano}. Early resonant metasurfaces, consisting of dielectric layers and metal films, exhibit relatively broad resonances and low transmission in the optical range due to high inherent energy dissipation of metal \cite{khurgin2015deal}. In recent years, all-dielectric metasurfaces with a high refractive index become an alternative choice in nanophotonics, due to their low dissipative losses, compatibility with semiconductor fabrication, and capabilities to trap the optical modes within dielectric nanostructures compared with their metallic counterparts \cite{kuznetsov2016optically,jahani2016all,huang2013general,liu2016q}. 

Bound states in the continuum (BICs) are states lying inside the continuum while remaining perfectly localized without radiation \cite{hsu2016bound,hsu2013,bogdanov2019bound,kodigala2017lasing,Nnano2018lasing,carletti2018giant,jin2019topologically,azzam2018formation}, which can facilitate the excitation of resonance in dielectric metasurfaces. Based on their mechanisms, BICs are divided into symmetry-protected BICs and accidental BICs \cite{koshelev2019nonradiating}. Symmetry-protected BICs can be perturbed through oblique incidence or symmetry breaking of nanostructures via building the radiation channel between eigenmodes and the free space, and thus exciting quasi-BICs \cite{cong2019symmetry,lee2012observation,liu2017high,multipolar2019,koshelev2018asymmetric,li2019symmetry,liu2019high,kupriianov2019metasurface,mikheeva2019photosensitive,koshelev2019nonlinear,xu2019dynamic,he2018toroidal}. Asymmetric Fano profiles with high contrast of transmission produced by the constructive and destructive interference of this localized state and the continuum have been excited in various metasurfaces. 

Actively tunable Fano-resonant metasurfaces can be realized through assistance of liquid crystal \cite{shen2019liquid,parry2017active}, two-dimensional materials \cite{zhou2018tunable,liactive} and phase-change materials \cite{cao2016controlling,chu2016active}, and further enable multifunctional active metadevices. $\rm Ge_2Sb_2Te_5$ (GST), one of phase-change materials, has received widespread attention recently attributing to its quick response, good stability, large number of switching cycles and significant refractive index contrast at amorphous and crystalline states \cite{wang2016optically,li2019active,tian2019active,qu2017dynamic,qu2017dynamic,karvounis2016all}. The transition between states of GST can be induced by thermal, electrical and optical means. Nowadays, there have been several mature methods to selectively modify phase states of GST on metasurfaces, such as conductive-AFM and laser-direct writing \cite{tseng2012fabrication,pandian2007nanoscale,chu2016active}. Combining the tunability of GST and high contrast of transmission in Fano-resonant dielectric metasurfaces, the transmitted power at a specific wavelength can be tuned dramatically. Therefore, they hold great potentials in reconfigurable real-time imaging.

In our work, we investigate the imaging function of Fano-resonant dielectric metasurface supporting quasi-BIC. Firstly, we study the excitation of quasi-BIC by introducing in-plane asymmetry in highly symmetric Si nanodisks. The far-field radiation and near-field dis-tribution are analyzed to make a deep discussion. Then we discuss the tunability of the resonance through in-troducing a GST film to the asymmetric metasurface. Finally, after changing amorphous GST ($a$-GST) to crystal-line GST ($c$-GST) at specific areas of a 40$\times$40 array, desired patterns can be encoded in the transmitted light and a reconfigurable imaging can be realized at 1553 nm wavelength. This active metasurface holds great potentials in fast spatial modulations. 

\section{\label{sec:2}Quasi-BIC supported by asymmetric nanodisks}
\subsection*{\rm \bf A. Characteristics of the excited Fano resonance.}

A perturbation that breaks the in-plane inversion symmetry of a structure can transform a symmetry-protected BIC into a quasi-BIC. As shown in Fig.~\ref{fig:1}(a), cutting a circular notch at the edge of the nanodisk introduces symmetry breaking. The center distance between the nanodisk and the air-hole is $x_{\Delta r_0}$=200 nm and the radius of air-hole is $r_{2}$=160 nm. The nanodisk with a height of $h$=220 nm and a radius of $r_1$=290 nm lies on a $\rm SiO_2$ substrate with a constant lattice of $P$= 1 $\mu$m. Finite-element method (COMSOL Multiphysics) and finite-difference-time-domain method (FDTD Solutions) are implemented to analyze the optical properties of metasurfaces. A $y$-polarized light is normally incident on metasurfaces along the $z$ direction. The periodical boundary conditions are set in the $x$ and $y$ directions and perfectly matched layers are set in the $z$ direction. The dielectric constants of Si and $\rm SiO_2$ are extracted from Palik Handbook \cite{palik1998handbook}. The calculated transmission spectrum is illustrated in Fig.~\ref{fig:1}(b), manifesting a sharp asymmetric Fano resonance with a maximum transmission of 1 and a minimum one of 0. The actual resonant wavelength 1561 nm locates exactly at half the distance between the maximum and minimum. Here we analyze the formation of Fano resonance in detail. After cutting a notch at the edge of the nanodisk, the electric dipole along the $y$ direction ($P_y$) produced by the incident polarization will exhibit slightly different dipole strengths in the notch part and the opposite side. These asymmetric two dipoles will generate a $z$-directed magnetic field near the center of the nanodisk, giving rise to the $z$-polarized magnetic dipole ($M_z$) . Finally, the broken symmetry induces interference between the in-plane electric dipole $P_y$ and the magnetic dipole $M_z$, leading to the observed Fano resonance.

The simulated electric field distributions at the resonant wavelength 1561 nm are shown in Fig.~\ref{fig:1}(c), with monitors placed at $z$=110 nm, $y$=0, $x$=-100 nm to detect field in the nanostructure. The polarization of the electric field is anti-parallel at opposite sides of the breaking nanodisk induced by the varied effective refractive index of silicon metasurface. This results in a distinct circular displacement current inside the nanodisk as represented by the blue arrows in the $x$-$y$ plane, revealing that the energy within the metasurface is strongly trapped by the magnetic dipole moment oscillation along the $z$ axis illustrated by red arrows in the center of nanodisk. In order to study the excitation of this Fano resonance, the electromagnetic multipole expansion under the Cartesian coordinate is performed, with the calculated scattered power shown in Fig.~\ref{fig:1}(d). At the resonance region, $M_z$ component makes a pronounced contribution, while other multipoles are dramatically suppressed. For the off-resonance region, $P_y$ component dominates. These results validate the above discussion of Fano excitation. 
\begin{figure}[htbp]
	\centering
	\includegraphics[scale=0.44]{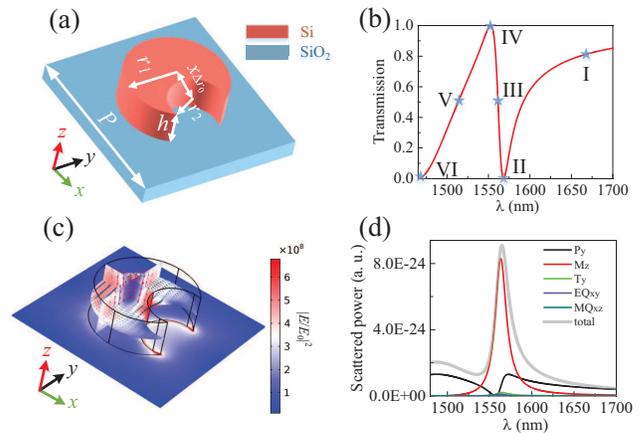}
	\caption{\label{fig:1} (a) Geometry of a unit cell composed of an asymmetric Si nanodisk after cutting a circular notch from the edge. (b) Simulated transmission spectrum of the metasurface. (c) Electric field distributions at different cross sections in the nanodisk at the resonant wavelength. Blue arrows and red arrows represent directions of the displacement current and magnetic field, respectively. (d) The calculated scattered power of dominant multipole components.}
	\label{fig:1}
\end{figure}

\subsection*{\rm \bf B. Far-field radiation patterns.}
We select six typical wavelengths near the Fano resonance as denoted by blue pentagrams and Roman numbers in Fig.~\ref{fig:1}(b) to compare far-field radiation patterns. Figure~\ref{fig:2} illustrates the far-field radiation patterns at $\lambda_{I}$=1671 nm, $\lambda_{II}$=1568 nm, $\lambda_{III}$=1561 nm, $\lambda_{IV}$=1553 nm, $\lambda_{V}$=1514 nm and $\lambda_{VI}$=1465 nm wavelengths. These patterns are drawn in COMSOL Multiphysics. For the resonant wavelength $\lambda_{III}$=1561 nm shown in Fig.~\ref{fig:2}(c), where magnetic dipole along the $z$ axis plays a predominant role, the far-field radiation pattern mainly expands in the $x$-$y$ plane. A noticeable contribution from electric dipole moment along the $y$-axis breaks the symmetry of scattering along the $x$ and $y$ direction, resulting in a stronger side-scattering directed along the polarization direction of the electric field, and also produces a weak power leakage along the $z$ axis. It is noteworthy that the far-field radiation power reaches the maximum at this actual resonant wavelength, as verified in the multipole expansion of Fig.~\ref{fig:1}(d). For neighboring wavelengths $\lambda_{II}$=1568 nm and $\lambda_{IV}$=1553 nm shown in Figs.~\ref{fig:2}(b) and~\ref{fig:2}(d), the contribution from originally dominant $M_z$ decreases, and its scattered power becomes comparable with that of $P_y$. Therefore, $P_y$ manifests itself in the far-field radiation by perturbing the symmetry of radiation pattern at the resonance point and producing more leakage at the $z$ direction. The radiation power is cut down as well. As the wavelength further increasing from the resonant position, reaching $\lambda_{I}$=1671 nm as illustrated in Fig.~\ref{fig:2}(a), the interaction of several multipoles (especially electric dipole and magnetic dipole) leads to a huge difference between the forward scattering and backward scattering. Contributions from these non-resonant multipoles give rise to a further suppression of total scattering as well. When reducing the wavelength to the off-resonance region, corresponding to $\lambda_{V}$=1514 nm drawn in Fig.~\ref{fig:2}(e), the scattered power from $M_z$ dramatically decreases, then $P_y$ becomes the most contributive component. Two radiation lobes along the $x$ axis emerge and the forward scattering exhibits an enhancement. While further deviating from the resonance as depicted in Fig.~\ref{fig:2}(f) at $\lambda_{VI}$=1465 nm, the contribution from magnetic dipole becomes negligible, thus the enhancement of forward scattering nearly disappears. 
\begin{figure}[t]
	\centering
	\includegraphics[scale=0.52]{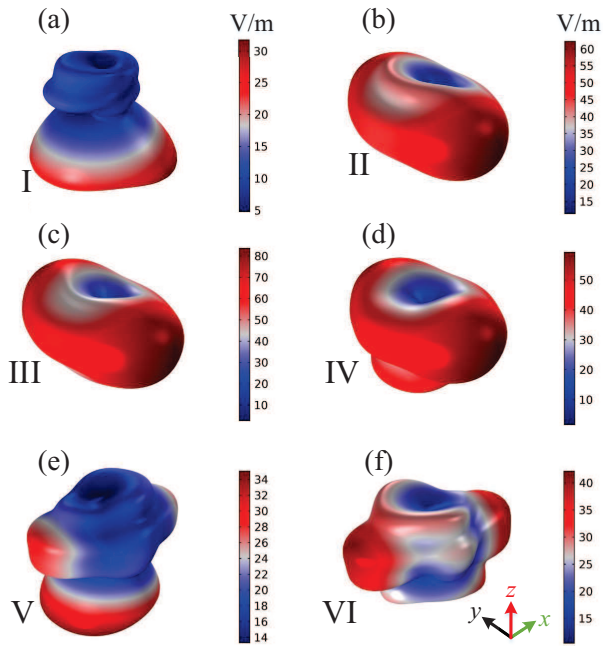}
	\caption{\label{fig:2} Radiation patterns of metasurface composed of asymmetric nanodisks at different wavelengths. (a) $\lambda_{I}$=1671 nm; (b) $\lambda_{II}$=1568 nm; (c) $\lambda_{III}$=1561 nm; (d) $\lambda_{IV}$=1553 nm; (e) $\lambda_{V}$=1514 nm; (f) $\lambda_{VI}$=1465 nm.}
	\label{fig:2}
\end{figure}

\section{\label{sec:3}Imaging through a GST-assisted Fano-resonant Metasurface}
In order to realize a reconfigurable imaging using the Fano-resonant metasurface, we introduce a GST film on the nanosdisk to manipulate the transmitted power at a specific wavelength, which is caused by the shift of resonance when GST transitions from $a$-GST to $c$-GST. The tunable characteristics of Fano resonance and the performance of Fano resonance-based imaging are demonstrated and discussed below.   

\subsection*{\rm \bf A. Active Fano resonance in the Si/GST metasurface}
Herein, we add a GST film with a thickness of 20 nm on the above metasurface with all other parameters remaining the same as shown in Fig.~\ref{fig:3}(a). The optical properties of GST are extracted from \cite{karvounis2016all}. At the same simulation configurations, the calculated transmissions are illustrated in Fig.~\ref{fig:3}(b). When adding an $a$-GST layer on the metasurface, the Fano resonance exhibits a slight red-shift compared to the transmission without GST, which is due to the increased effective refractive index of structure introduced by the GST film. The slightly reduced extinction ratio and broadened resonance are ascribed to the additional absorption introduced by $a$-GST \cite{qu2017dynamic}. After changing $a$-GST to $c$-GST, the resonance wavelength increases about 50 nm due to the much larger refractive index of $c$-GST. In addition, the extinction ratio of the resonance suffers from an obvious reduction, providing by the growing absorption of $c$-GST given by its larger extinction coefficient. 

To make the numerical calculation more available, we focus our investigation of imaging on a metasurface composed of 40$\times$40 nanodisks. By setting the boundary conditions at the $x$ and $y$ directions as perfectly matched layers, we calculate the transmissions of these 40$\times$40 arrays (Fig.~\ref{fig:3}(c)) at $a$-GST and $c$-GST states shown in Fig.~\ref{fig:3}(d). An obvious reduction of extinction ratio is observed compared with Fig.~\ref{fig:3}(b), which can be ascribed to the array effect \cite{fedotov2010spectral,yang2014all,campione2016broken}. The resonance shown in Fig.~\ref{fig:3}(b) is excited by a unit cell with periodic boundaries at both $x$ and $y$ directions, which corresponds to an infinite array. As magnetic dipoles along the $z$ axis supported by each unit coherently oscillate perpendicular to the array plane, they cancel each other and the resonance only scatters at edges of the metasurface. When the array becomes smaller as the 40$\times$40 array, the scattering losses grow and the resonance becomes weaker due to scattering losses of magnetic dipole radiation. It is noted that the resonance at the $a$-GST state is influenced more strongly by the array effect compared with that at the $c$-GST state, whose reason lies in the sharper resonance hold by the $a$-GST state.  

\begin{figure}[htbp]
	\centering
	\includegraphics[scale=0.44]{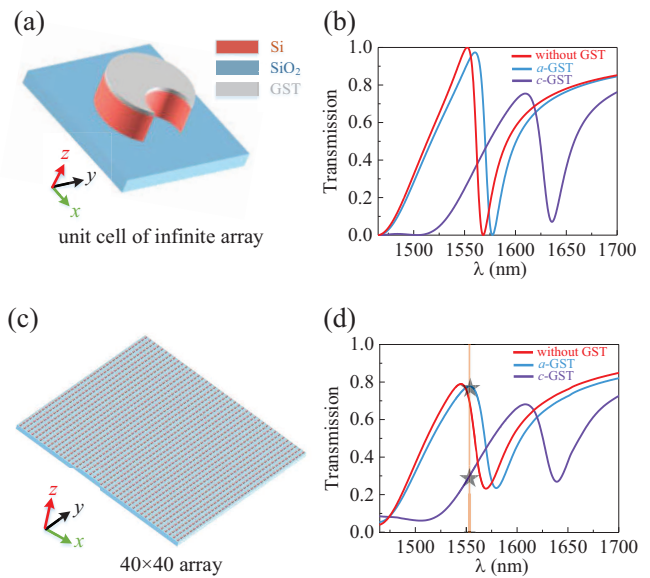}
	\caption{\label{fig:3} (a) Geometry of a unit cell of an infinite array with Si/GST asymmetric nanodisks. (b) Simulated transmission spectra of the infinite array. (c) Geometry of a 40$\times$40 array with Si/GST asymmetric nanodisks. (b) Simulated transmission spectra of the 40$\times$40 array.}
	\label{fig:3}
\end{figure}

\subsection*{\rm \bf B. Reconfigurable imaging}

We choose an array consisted of 40$\times$40 nanodisk units as shown in Fig.~\ref{fig:3}(c) to balance the imaging performance and simulation capacity. The operation wavelength is set at 1553 nm (denoted by the yellow line in Fig.~\ref{fig:3}(d)), the transmission maximum at the $a$-GST state. Thus when GST transitions from the amorphous state to the crystalline state, the corresponding transmission changes from 0.78 to 0.29 (marked by black pentagrams). We use letters "GZMU" as target images to demonstrate the imaging function of our active metasurface. First, each image of letters is discretized and then represented by a binary matrix with a size of 40$\times$40, with pixels corresponding to the letter part valued 1 and other pixels valued 0. Then phase states of GST on units of the metasurface are altered based on the target image matrix. This selective modification of GST phase states can be achieved in experiment through laser-direct writing, conductive-AFM and so on. For matrix pixels with values of 0, the corresponding $a$-GST is changed to $c$-GST, thus the transmitted power from them becomes weaker compared with the other areas covered by $a$-GST, which manifests the desired pattern on the transmitted region. We place a field distribution monitor 10 $\mu$m away from the metasurface in the transmitted region to capture and record power distribution of images, which is defined as the square of electric field abstract value. All these target and transmitted images are shown in Fig.~\ref{fig:4}, with each image normalized to itself. When the incident polarization relative to target images shown in this figure is along the perpendicular direction, the transmitted images are shown in Figs.~\ref{fig:4}(e)-~\ref{fig:4}(h). All targets are clearly imaged, with power at corresponding positions of letter pixels significantly enhanced with respect to the background part. At edges of imaged letters, the power is relatively weak with no obvious boundaries with the background like that in targets. That is because the destructive interference between $a$-GST pixels and $c$-GST pixels cuts down the power at their interface. There are also alternating bright and dark stripes in the imaging region, which can be ascribed to the constructive and destructive interference of light scattered from different units. Imaging performances are also influenced by shapes of target letters. At crossing points of features in letters, where more $a$-GST units are included, the collected power is stronger. It is resulted from the enhanced constructive interference in these larger areas of $a$-GST units with same amplitude and phase responses. Features along the polarization direction are more distinguishable with more power accumulated, which indicates an anisotropic response of imaging. While for the letter "M", its pixel distribution may play an important role for the relatively weak imaging. Because there exists diffraction at the boundary of metasurface, the profile of transmitted light will be redistributed due to the interference between this diffraction effect and the light steering effects supported by unit cells, tailoring the relative intensities in the detected position. Given that the left and right sides in "M" approach to the edge of metasurface, it may be more sensitive to the diffraction effect and thus be perturbed more obviously by the redistribution phenomenon.

\begin{figure}[t]
	\centering
	\includegraphics[scale=0.4]{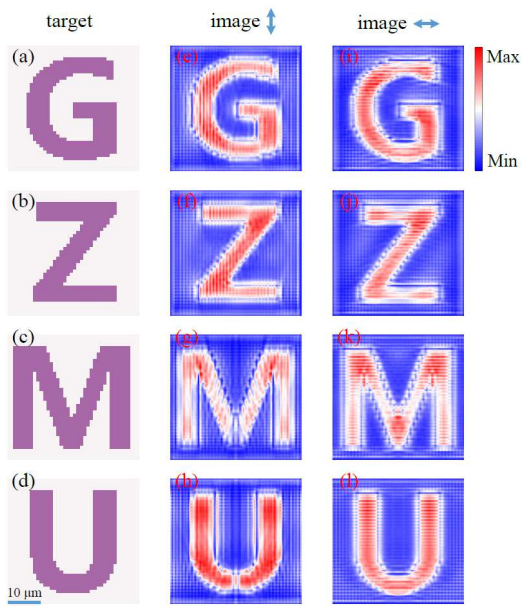}
	\caption{\label{fig:4} (a)-(d) Target images of letters used to perform the imaging function using metasurfaces. (e)-(h) Produced transmitted images when the incident light polarized along the perpendicular direction with respect to targets. (i)-(l) Produced transmitted images when the incident light polarized along the horizontal direction with respect to targets. Sizes of images are all 40 $\mu$m$\times$40 $\mu$m, with the scale bar marked by the blue line.}
	\label{fig:4}
\end{figure}

It is noted that our designed metasurface is polarization-dependent, where the Fano resonance can only be excited by incident polarization perpendicular to the symmetry axis of the nanodisk in the $x-y$ plane. So this imaging function can only be performed under incident light polarized along the $y$ axis. Through rotating target images while keeping the incident polarization unchanged, we also investigate the response of metasurface when the incident polarization is along the horizontal direction with respect to targets, with the results depicted in Figs.~\ref{fig:4}(i)-~\ref{fig:4}(l). Transmitted image patterns can still be observed clearly. For the target letter "U", the produced image is not as strong and distinct as that for another polarization, whose reason may lie in the weakened power along the perpendicular direction when the incident light polarized along the horizontal direction.

The imaging function of this metasurface is tunable and reconfigurable given that GST can transition from amorphous to crystalline states for a large number of cycles just in a few nanoseconds. So through selectively modify the phase states of GST between $a$-GST and $c$-GST to form different patterns, this metasurface can realize real-time reconfigurable imaging with high speed.

\section{\label{sec:4}Conclusions}
In summary, we investigate the imaging function of a dielectric metasurface supporting a strong Fano resonance. First, we demonstrate that this Fano resonance is governed by the excitation of quasi-BIC, through breaking the in-plane symmetry of nanodisks to perturb the symmetry-protected BIC. Then near-field distributions and far-field multipole contributions are carried out to validate the dominant role of magnetic dipole along the $z$ axis in the excitation of Fano resonance. Radiation patterns in the far-field are depicted for several wavelengths near the resonance to further analyze resonant characteristics. Next, after adding a GST film on the metasurface, we discuss its imaging function in detail. We begin with analyzing transmission properties of this Si/GST metasurface. Then we use a 40$\times$40 array to produce transmitted images of targets via selective modifying phase states of GST at different units from $a$-GST to $c$-GST, which well reconstruct corresponding targets. Our work may open up a route for metasurface-based imaging and spatial light modulation with good performance.

\begin{acknowledgments}
This work is supported by the Natural Science Research Project of Guizhou Minzu University (GZMU[2019]YB22).
\end{acknowledgments}

\bibliography{manuscript}

\begin{thebibliography}{45}%
\makeatletter
\providecommand \@ifxundefined [1]{%
 \@ifx{#1\undefined}
}%
\providecommand \@ifnum [1]{%
 \ifnum #1\expandafter \@firstoftwo
 \else \expandafter \@secondoftwo
 \fi
}%
\providecommand \@ifx [1]{%
 \ifx #1\expandafter \@firstoftwo
 \else \expandafter \@secondoftwo
 \fi
}%
\providecommand \natexlab [1]{#1}%
\providecommand \enquote  [1]{``#1''}%
\providecommand \bibnamefont  [1]{#1}%
\providecommand \bibfnamefont [1]{#1}%
\providecommand \citenamefont [1]{#1}%
\providecommand \href@noop [0]{\@secondoftwo}%
\providecommand \href [0]{\begingroup \@sanitize@url \@href}%
\providecommand \@href[1]{\@@startlink{#1}\@@href}%
\providecommand \@@href[1]{\endgroup#1\@@endlink}%
\providecommand \@sanitize@url [0]{\catcode `\\12\catcode `\$12\catcode
  `\&12\catcode `\#12\catcode `\^12\catcode `\_12\catcode `\%12\relax}%
\providecommand \@@startlink[1]{}%
\providecommand \@@endlink[0]{}%
\providecommand \url  [0]{\begingroup\@sanitize@url \@url }%
\providecommand \@url [1]{\endgroup\@href {#1}{\urlprefix }}%
\providecommand \urlprefix  [0]{URL }%
\providecommand \Eprint [0]{\href }%
\providecommand \doibase [0]{http://dx.doi.org/}%
\providecommand \selectlanguage [0]{\@gobble}%
\providecommand \bibinfo  [0]{\@secondoftwo}%
\providecommand \bibfield  [0]{\@secondoftwo}%
\providecommand \translation [1]{[#1]}%
\providecommand \BibitemOpen [0]{}%
\providecommand \bibitemStop [0]{}%
\providecommand \bibitemNoStop [0]{.\EOS\space}%
\providecommand \EOS [0]{\spacefactor3000\relax}%
\providecommand \BibitemShut  [1]{\csname bibitem#1\endcsname}%
\let\auto@bib@innerbib\@empty
\bibitem [{\citenamefont {Miroshnichenko}\ \emph {et~al.}(2010)\citenamefont
  {Miroshnichenko}, \citenamefont {Flach},\ and\ \citenamefont
  {Kivshar}}]{miroshnichenko2010fano}%
  \BibitemOpen
  \bibfield  {author} {\bibinfo {author} {\bibfnamefont {A.~E.}\ \bibnamefont
  {Miroshnichenko}}, \bibinfo {author} {\bibfnamefont {S.}~\bibnamefont
  {Flach}}, \ and\ \bibinfo {author} {\bibfnamefont {Y.~S.}\ \bibnamefont
  {Kivshar}},\ }\href@noop {} {\bibfield  {journal} {\bibinfo  {journal}
  {Reviews of Modern Physics}\ }\textbf {\bibinfo {volume} {82}},\ \bibinfo
  {pages} {2257} (\bibinfo {year} {2010})}\BibitemShut {NoStop}%
\bibitem [{\citenamefont {Limonov}\ \emph {et~al.}(2017)\citenamefont
  {Limonov}, \citenamefont {Rybin}, \citenamefont {Poddubny},\ and\
  \citenamefont {Kivshar}}]{limonov2017fano}%
  \BibitemOpen
  \bibfield  {author} {\bibinfo {author} {\bibfnamefont {M.~F.}\ \bibnamefont
  {Limonov}}, \bibinfo {author} {\bibfnamefont {M.~V.}\ \bibnamefont {Rybin}},
  \bibinfo {author} {\bibfnamefont {A.~N.}\ \bibnamefont {Poddubny}}, \ and\
  \bibinfo {author} {\bibfnamefont {Y.~S.}\ \bibnamefont {Kivshar}},\
  }\href@noop {} {\bibfield  {journal} {\bibinfo  {journal} {Nature Photonics}\
  }\textbf {\bibinfo {volume} {11}},\ \bibinfo {pages} {543} (\bibinfo {year}
  {2017})}\BibitemShut {NoStop}%
\bibitem [{\citenamefont {Khurgin}(2015)}]{khurgin2015deal}%
  \BibitemOpen
  \bibfield  {author} {\bibinfo {author} {\bibfnamefont {J.~B.}\ \bibnamefont
  {Khurgin}},\ }\href@noop {} {\bibfield  {journal} {\bibinfo  {journal}
  {Nature nanotechnology}\ }\textbf {\bibinfo {volume} {10}},\ \bibinfo {pages}
  {2} (\bibinfo {year} {2015})}\BibitemShut {NoStop}%
\bibitem [{\citenamefont {Kuznetsov}\ \emph {et~al.}(2016)\citenamefont
  {Kuznetsov}, \citenamefont {Miroshnichenko}, \citenamefont {Brongersma},
  \citenamefont {Kivshar},\ and\ \citenamefont
  {Luk’yanchuk}}]{kuznetsov2016optically}%
  \BibitemOpen
  \bibfield  {author} {\bibinfo {author} {\bibfnamefont {A.~I.}\ \bibnamefont
  {Kuznetsov}}, \bibinfo {author} {\bibfnamefont {A.~E.}\ \bibnamefont
  {Miroshnichenko}}, \bibinfo {author} {\bibfnamefont {M.~L.}\ \bibnamefont
  {Brongersma}}, \bibinfo {author} {\bibfnamefont {Y.~S.}\ \bibnamefont
  {Kivshar}}, \ and\ \bibinfo {author} {\bibfnamefont {B.}~\bibnamefont
  {Luk’yanchuk}},\ }\href@noop {} {\bibfield  {journal} {\bibinfo  {journal}
  {Science}\ }\textbf {\bibinfo {volume} {354}},\ \bibinfo {pages} {aag2472}
  (\bibinfo {year} {2016})}\BibitemShut {NoStop}%
\bibitem [{\citenamefont {Jahani}\ and\ \citenamefont
  {Jacob}(2016)}]{jahani2016all}%
  \BibitemOpen
  \bibfield  {author} {\bibinfo {author} {\bibfnamefont {S.}~\bibnamefont
  {Jahani}}\ and\ \bibinfo {author} {\bibfnamefont {Z.}~\bibnamefont {Jacob}},\
  }\href@noop {} {\bibfield  {journal} {\bibinfo  {journal} {Nature
  nanotechnology}\ }\textbf {\bibinfo {volume} {11}},\ \bibinfo {pages} {23}
  (\bibinfo {year} {2016})}\BibitemShut {NoStop}%
\bibitem [{\citenamefont {Huang}\ \emph {et~al.}(2013)\citenamefont {Huang},
  \citenamefont {Yu},\ and\ \citenamefont {Cao}}]{huang2013general}%
  \BibitemOpen
  \bibfield  {author} {\bibinfo {author} {\bibfnamefont {L.}~\bibnamefont
  {Huang}}, \bibinfo {author} {\bibfnamefont {Y.}~\bibnamefont {Yu}}, \ and\
  \bibinfo {author} {\bibfnamefont {L.}~\bibnamefont {Cao}},\ }\href@noop {}
  {\bibfield  {journal} {\bibinfo  {journal} {Nano letters}\ }\textbf {\bibinfo
  {volume} {13}},\ \bibinfo {pages} {3559} (\bibinfo {year}
  {2013})}\BibitemShut {NoStop}%
\bibitem [{\citenamefont {Liu}\ \emph {et~al.}(2016)\citenamefont {Liu},
  \citenamefont {Miroshnichenko},\ and\ \citenamefont {Kivshar}}]{liu2016q}%
  \BibitemOpen
  \bibfield  {author} {\bibinfo {author} {\bibfnamefont {W.}~\bibnamefont
  {Liu}}, \bibinfo {author} {\bibfnamefont {A.~E.}\ \bibnamefont
  {Miroshnichenko}}, \ and\ \bibinfo {author} {\bibfnamefont {Y.~S.}\
  \bibnamefont {Kivshar}},\ }\href@noop {} {\bibfield  {journal} {\bibinfo
  {journal} {Physical Review B}\ }\textbf {\bibinfo {volume} {94}},\ \bibinfo
  {pages} {195436} (\bibinfo {year} {2016})}\BibitemShut {NoStop}%
\bibitem [{\citenamefont {Hsu}\ \emph {et~al.}(2016)\citenamefont {Hsu},
  \citenamefont {Zhen}, \citenamefont {Stone}, \citenamefont {Joannopoulos},\
  and\ \citenamefont {Solja{\v{c}}i{\'c}}}]{hsu2016bound}%
  \BibitemOpen
  \bibfield  {author} {\bibinfo {author} {\bibfnamefont {C.~W.}\ \bibnamefont
  {Hsu}}, \bibinfo {author} {\bibfnamefont {B.}~\bibnamefont {Zhen}}, \bibinfo
  {author} {\bibfnamefont {A.~D.}\ \bibnamefont {Stone}}, \bibinfo {author}
  {\bibfnamefont {J.~D.}\ \bibnamefont {Joannopoulos}}, \ and\ \bibinfo
  {author} {\bibfnamefont {M.}~\bibnamefont {Solja{\v{c}}i{\'c}}},\ }\href@noop
  {} {\bibfield  {journal} {\bibinfo  {journal} {Nature Reviews Materials}\
  }\textbf {\bibinfo {volume} {1}},\ \bibinfo {pages} {16048} (\bibinfo {year}
  {2016})}\BibitemShut {NoStop}%
\bibitem [{\citenamefont {Hsu}\ \emph {et~al.}(2013)\citenamefont {Hsu},
  \citenamefont {Zhen}, \citenamefont {Lee}, \citenamefont {Chua},
  \citenamefont {Johnson}, \citenamefont {Joannopoulos},\ and\ \citenamefont
  {Solja{\v{c}}i{\'c}}}]{hsu2013}%
  \BibitemOpen
  \bibfield  {author} {\bibinfo {author} {\bibfnamefont {C.~W.}\ \bibnamefont
  {Hsu}}, \bibinfo {author} {\bibfnamefont {B.}~\bibnamefont {Zhen}}, \bibinfo
  {author} {\bibfnamefont {J.}~\bibnamefont {Lee}}, \bibinfo {author}
  {\bibfnamefont {S.-L.}\ \bibnamefont {Chua}}, \bibinfo {author}
  {\bibfnamefont {S.~G.}\ \bibnamefont {Johnson}}, \bibinfo {author}
  {\bibfnamefont {J.~D.}\ \bibnamefont {Joannopoulos}}, \ and\ \bibinfo
  {author} {\bibfnamefont {M.}~\bibnamefont {Solja{\v{c}}i{\'c}}},\ }\href@noop
  {} {\bibfield  {journal} {\bibinfo  {journal} {Nature}\ }\textbf {\bibinfo
  {volume} {499}},\ \bibinfo {pages} {188} (\bibinfo {year}
  {2013})}\BibitemShut {NoStop}%
\bibitem [{\citenamefont {Bogdanov}\ \emph {et~al.}(2019)\citenamefont
  {Bogdanov}, \citenamefont {Koshelev}, \citenamefont {Kapitanova},
  \citenamefont {Rybin}, \citenamefont {Gladyshev}, \citenamefont {Sadrieva},
  \citenamefont {Samusev}, \citenamefont {Kivshar},\ and\ \citenamefont
  {Limonov}}]{bogdanov2019bound}%
  \BibitemOpen
  \bibfield  {author} {\bibinfo {author} {\bibfnamefont {A.~A.}\ \bibnamefont
  {Bogdanov}}, \bibinfo {author} {\bibfnamefont {K.~L.}\ \bibnamefont
  {Koshelev}}, \bibinfo {author} {\bibfnamefont {P.~V.}\ \bibnamefont
  {Kapitanova}}, \bibinfo {author} {\bibfnamefont {M.~V.}\ \bibnamefont
  {Rybin}}, \bibinfo {author} {\bibfnamefont {S.~A.}\ \bibnamefont
  {Gladyshev}}, \bibinfo {author} {\bibfnamefont {Z.~F.}\ \bibnamefont
  {Sadrieva}}, \bibinfo {author} {\bibfnamefont {K.~B.}\ \bibnamefont
  {Samusev}}, \bibinfo {author} {\bibfnamefont {Y.~S.}\ \bibnamefont
  {Kivshar}}, \ and\ \bibinfo {author} {\bibfnamefont {M.~F.}\ \bibnamefont
  {Limonov}},\ }\href@noop {} {\bibfield  {journal} {\bibinfo  {journal}
  {Advanced Photonics}\ }\textbf {\bibinfo {volume} {1}},\ \bibinfo {pages}
  {016001} (\bibinfo {year} {2019})}\BibitemShut {NoStop}%
\bibitem [{\citenamefont {Kodigala}\ \emph {et~al.}(2017)\citenamefont
  {Kodigala}, \citenamefont {Lepetit}, \citenamefont {Gu}, \citenamefont
  {Bahari}, \citenamefont {Fainman},\ and\ \citenamefont
  {Kant{\'e}}}]{kodigala2017lasing}%
  \BibitemOpen
  \bibfield  {author} {\bibinfo {author} {\bibfnamefont {A.}~\bibnamefont
  {Kodigala}}, \bibinfo {author} {\bibfnamefont {T.}~\bibnamefont {Lepetit}},
  \bibinfo {author} {\bibfnamefont {Q.}~\bibnamefont {Gu}}, \bibinfo {author}
  {\bibfnamefont {B.}~\bibnamefont {Bahari}}, \bibinfo {author} {\bibfnamefont
  {Y.}~\bibnamefont {Fainman}}, \ and\ \bibinfo {author} {\bibfnamefont
  {B.}~\bibnamefont {Kant{\'e}}},\ }\href@noop {} {\bibfield  {journal}
  {\bibinfo  {journal} {Nature}\ }\textbf {\bibinfo {volume} {541}},\ \bibinfo
  {pages} {196} (\bibinfo {year} {2017})}\BibitemShut {NoStop}%
\bibitem [{\citenamefont {Ha}\ \emph {et~al.}(2018)\citenamefont {Ha},
  \citenamefont {Fu}, \citenamefont {Emani}, \citenamefont {Pan}, \citenamefont
  {Bakker}, \citenamefont {Paniagua-Dom{\'\i}nguez},\ and\ \citenamefont
  {Kuznetsov}}]{Nnano2018lasing}%
  \BibitemOpen
  \bibfield  {author} {\bibinfo {author} {\bibfnamefont {S.~T.}\ \bibnamefont
  {Ha}}, \bibinfo {author} {\bibfnamefont {Y.~H.}\ \bibnamefont {Fu}}, \bibinfo
  {author} {\bibfnamefont {N.~K.}\ \bibnamefont {Emani}}, \bibinfo {author}
  {\bibfnamefont {Z.}~\bibnamefont {Pan}}, \bibinfo {author} {\bibfnamefont
  {R.~M.}\ \bibnamefont {Bakker}}, \bibinfo {author} {\bibfnamefont
  {R.}~\bibnamefont {Paniagua-Dom{\'\i}nguez}}, \ and\ \bibinfo {author}
  {\bibfnamefont {A.~I.}\ \bibnamefont {Kuznetsov}},\ }\href@noop {} {\bibfield
   {journal} {\bibinfo  {journal} {Nat. Nanotechnology}\ }\textbf {\bibinfo
  {volume} {13}},\ \bibinfo {pages} {1042} (\bibinfo {year}
  {2018})}\BibitemShut {NoStop}%
\bibitem [{\citenamefont {Carletti}\ \emph {et~al.}(2018)\citenamefont
  {Carletti}, \citenamefont {Koshelev}, \citenamefont {De~Angelis},\ and\
  \citenamefont {Kivshar}}]{carletti2018giant}%
  \BibitemOpen
  \bibfield  {author} {\bibinfo {author} {\bibfnamefont {L.}~\bibnamefont
  {Carletti}}, \bibinfo {author} {\bibfnamefont {K.}~\bibnamefont {Koshelev}},
  \bibinfo {author} {\bibfnamefont {C.}~\bibnamefont {De~Angelis}}, \ and\
  \bibinfo {author} {\bibfnamefont {Y.}~\bibnamefont {Kivshar}},\ }\href@noop
  {} {\bibfield  {journal} {\bibinfo  {journal} {Physical review letters}\
  }\textbf {\bibinfo {volume} {121}},\ \bibinfo {pages} {033903} (\bibinfo
  {year} {2018})}\BibitemShut {NoStop}%
\bibitem [{\citenamefont {Jin}\ \emph {et~al.}(2019)\citenamefont {Jin},
  \citenamefont {Yin}, \citenamefont {Ni}, \citenamefont {Solja{\v{c}}i{\'c}},
  \citenamefont {Zhen},\ and\ \citenamefont {Peng}}]{jin2019topologically}%
  \BibitemOpen
  \bibfield  {author} {\bibinfo {author} {\bibfnamefont {J.}~\bibnamefont
  {Jin}}, \bibinfo {author} {\bibfnamefont {X.}~\bibnamefont {Yin}}, \bibinfo
  {author} {\bibfnamefont {L.}~\bibnamefont {Ni}}, \bibinfo {author}
  {\bibfnamefont {M.}~\bibnamefont {Solja{\v{c}}i{\'c}}}, \bibinfo {author}
  {\bibfnamefont {B.}~\bibnamefont {Zhen}}, \ and\ \bibinfo {author}
  {\bibfnamefont {C.}~\bibnamefont {Peng}},\ }\href@noop {} {\bibfield
  {journal} {\bibinfo  {journal} {Nature}\ }\textbf {\bibinfo {volume} {574}},\
  \bibinfo {pages} {501} (\bibinfo {year} {2019})}\BibitemShut {NoStop}%
\bibitem [{\citenamefont {Azzam}\ \emph {et~al.}(2018)\citenamefont {Azzam},
  \citenamefont {Shalaev}, \citenamefont {Boltasseva},\ and\ \citenamefont
  {Kildishev}}]{azzam2018formation}%
  \BibitemOpen
  \bibfield  {author} {\bibinfo {author} {\bibfnamefont {S.~I.}\ \bibnamefont
  {Azzam}}, \bibinfo {author} {\bibfnamefont {V.~M.}\ \bibnamefont {Shalaev}},
  \bibinfo {author} {\bibfnamefont {A.}~\bibnamefont {Boltasseva}}, \ and\
  \bibinfo {author} {\bibfnamefont {A.~V.}\ \bibnamefont {Kildishev}},\
  }\href@noop {} {\bibfield  {journal} {\bibinfo  {journal} {Physical review
  letters}\ }\textbf {\bibinfo {volume} {121}},\ \bibinfo {pages} {253901}
  (\bibinfo {year} {2018})}\BibitemShut {NoStop}%
\bibitem [{\citenamefont {Koshelev}\ \emph
  {et~al.}(2019{\natexlab{a}})\citenamefont {Koshelev}, \citenamefont
  {Favraud}, \citenamefont {Bogdanov}, \citenamefont {Kivshar},\ and\
  \citenamefont {Fratalocchi}}]{koshelev2019nonradiating}%
  \BibitemOpen
  \bibfield  {author} {\bibinfo {author} {\bibfnamefont {K.}~\bibnamefont
  {Koshelev}}, \bibinfo {author} {\bibfnamefont {G.}~\bibnamefont {Favraud}},
  \bibinfo {author} {\bibfnamefont {A.}~\bibnamefont {Bogdanov}}, \bibinfo
  {author} {\bibfnamefont {Y.}~\bibnamefont {Kivshar}}, \ and\ \bibinfo
  {author} {\bibfnamefont {A.}~\bibnamefont {Fratalocchi}},\ }\href@noop {}
  {\bibfield  {journal} {\bibinfo  {journal} {Nanophotonics}\ }\textbf
  {\bibinfo {volume} {8}},\ \bibinfo {pages} {725} (\bibinfo {year}
  {2019}{\natexlab{a}})}\BibitemShut {NoStop}%
\bibitem [{\citenamefont {Cong}\ and\ \citenamefont
  {Singh}(2019)}]{cong2019symmetry}%
  \BibitemOpen
  \bibfield  {author} {\bibinfo {author} {\bibfnamefont {L.}~\bibnamefont
  {Cong}}\ and\ \bibinfo {author} {\bibfnamefont {R.}~\bibnamefont {Singh}},\
  }\href@noop {} {\bibfield  {journal} {\bibinfo  {journal} {Advanced Optical
  Materials}\ ,\ \bibinfo {pages} {1900383}} (\bibinfo {year}
  {2019})}\BibitemShut {NoStop}%
\bibitem [{\citenamefont {Lee}\ \emph {et~al.}(2012)\citenamefont {Lee},
  \citenamefont {Zhen}, \citenamefont {Chua}, \citenamefont {Qiu},
  \citenamefont {Joannopoulos}, \citenamefont {Solja{\v{c}}i{\'c}},\ and\
  \citenamefont {Shapira}}]{lee2012observation}%
  \BibitemOpen
  \bibfield  {author} {\bibinfo {author} {\bibfnamefont {J.}~\bibnamefont
  {Lee}}, \bibinfo {author} {\bibfnamefont {B.}~\bibnamefont {Zhen}}, \bibinfo
  {author} {\bibfnamefont {S.-L.}\ \bibnamefont {Chua}}, \bibinfo {author}
  {\bibfnamefont {W.}~\bibnamefont {Qiu}}, \bibinfo {author} {\bibfnamefont
  {J.~D.}\ \bibnamefont {Joannopoulos}}, \bibinfo {author} {\bibfnamefont
  {M.}~\bibnamefont {Solja{\v{c}}i{\'c}}}, \ and\ \bibinfo {author}
  {\bibfnamefont {O.}~\bibnamefont {Shapira}},\ }\href@noop {} {\bibfield
  {journal} {\bibinfo  {journal} {Physical review letters}\ }\textbf {\bibinfo
  {volume} {109}},\ \bibinfo {pages} {067401} (\bibinfo {year}
  {2012})}\BibitemShut {NoStop}%
\bibitem [{\citenamefont {Liu}\ \emph {et~al.}(2017)\citenamefont {Liu},
  \citenamefont {Wang}, \citenamefont {Zhao}, \citenamefont {Zhou},\ and\
  \citenamefont {Sun}}]{liu2017high}%
  \BibitemOpen
  \bibfield  {author} {\bibinfo {author} {\bibfnamefont {Y.}~\bibnamefont
  {Liu}}, \bibinfo {author} {\bibfnamefont {S.}~\bibnamefont {Wang}}, \bibinfo
  {author} {\bibfnamefont {D.}~\bibnamefont {Zhao}}, \bibinfo {author}
  {\bibfnamefont {W.}~\bibnamefont {Zhou}}, \ and\ \bibinfo {author}
  {\bibfnamefont {Y.}~\bibnamefont {Sun}},\ }\href@noop {} {\bibfield
  {journal} {\bibinfo  {journal} {Optics express}\ }\textbf {\bibinfo {volume}
  {25}},\ \bibinfo {pages} {10536} (\bibinfo {year} {2017})}\BibitemShut
  {NoStop}%
\bibitem [{\citenamefont {Sadrieva}\ \emph {et~al.}(2019)\citenamefont
  {Sadrieva}, \citenamefont {Frizyuk}, \citenamefont {Petrov}, \citenamefont
  {Kivshar},\ and\ \citenamefont {Bogdanov}}]{multipolar2019}%
  \BibitemOpen
  \bibfield  {author} {\bibinfo {author} {\bibfnamefont {Z.}~\bibnamefont
  {Sadrieva}}, \bibinfo {author} {\bibfnamefont {K.}~\bibnamefont {Frizyuk}},
  \bibinfo {author} {\bibfnamefont {M.}~\bibnamefont {Petrov}}, \bibinfo
  {author} {\bibfnamefont {Y.}~\bibnamefont {Kivshar}}, \ and\ \bibinfo
  {author} {\bibfnamefont {A.}~\bibnamefont {Bogdanov}},\ }\href@noop {}
  {\bibfield  {journal} {\bibinfo  {journal} {Phys. Rev. B}\ }\textbf {\bibinfo
  {volume} {100}},\ \bibinfo {pages} {115303} (\bibinfo {year}
  {2019})}\BibitemShut {NoStop}%
\bibitem [{\citenamefont {Koshelev}\ \emph {et~al.}(2018)\citenamefont
  {Koshelev}, \citenamefont {Lepeshov}, \citenamefont {Liu}, \citenamefont
  {Bogdanov},\ and\ \citenamefont {Kivshar}}]{koshelev2018asymmetric}%
  \BibitemOpen
  \bibfield  {author} {\bibinfo {author} {\bibfnamefont {K.}~\bibnamefont
  {Koshelev}}, \bibinfo {author} {\bibfnamefont {S.}~\bibnamefont {Lepeshov}},
  \bibinfo {author} {\bibfnamefont {M.}~\bibnamefont {Liu}}, \bibinfo {author}
  {\bibfnamefont {A.}~\bibnamefont {Bogdanov}}, \ and\ \bibinfo {author}
  {\bibfnamefont {Y.}~\bibnamefont {Kivshar}},\ }\href@noop {} {\bibfield
  {journal} {\bibinfo  {journal} {Physical review letters}\ }\textbf {\bibinfo
  {volume} {121}},\ \bibinfo {pages} {193903} (\bibinfo {year}
  {2018})}\BibitemShut {NoStop}%
\bibitem [{\citenamefont {Li}\ \emph {et~al.}(2019{\natexlab{a}})\citenamefont
  {Li}, \citenamefont {Zhou}, \citenamefont {Liu},\ and\ \citenamefont
  {Xiao}}]{li2019symmetry}%
  \BibitemOpen
  \bibfield  {author} {\bibinfo {author} {\bibfnamefont {S.}~\bibnamefont
  {Li}}, \bibinfo {author} {\bibfnamefont {C.}~\bibnamefont {Zhou}}, \bibinfo
  {author} {\bibfnamefont {T.}~\bibnamefont {Liu}}, \ and\ \bibinfo {author}
  {\bibfnamefont {S.}~\bibnamefont {Xiao}},\ }\href@noop {} {\bibfield
  {journal} {\bibinfo  {journal} {Physical Review A}\ }\textbf {\bibinfo
  {volume} {100}},\ \bibinfo {pages} {063803} (\bibinfo {year}
  {2019}{\natexlab{a}})}\BibitemShut {NoStop}%
\bibitem [{\citenamefont {Liu}\ \emph {et~al.}(2019)\citenamefont {Liu},
  \citenamefont {Xu}, \citenamefont {Lin}, \citenamefont {Xiang}, \citenamefont
  {Feng}, \citenamefont {Cao}, \citenamefont {Li}, \citenamefont {Lan},\ and\
  \citenamefont {Liu}}]{liu2019high}%
  \BibitemOpen
  \bibfield  {author} {\bibinfo {author} {\bibfnamefont {Z.}~\bibnamefont
  {Liu}}, \bibinfo {author} {\bibfnamefont {Y.}~\bibnamefont {Xu}}, \bibinfo
  {author} {\bibfnamefont {Y.}~\bibnamefont {Lin}}, \bibinfo {author}
  {\bibfnamefont {J.}~\bibnamefont {Xiang}}, \bibinfo {author} {\bibfnamefont
  {T.}~\bibnamefont {Feng}}, \bibinfo {author} {\bibfnamefont {Q.}~\bibnamefont
  {Cao}}, \bibinfo {author} {\bibfnamefont {J.}~\bibnamefont {Li}}, \bibinfo
  {author} {\bibfnamefont {S.}~\bibnamefont {Lan}}, \ and\ \bibinfo {author}
  {\bibfnamefont {J.}~\bibnamefont {Liu}},\ }\href@noop {} {\bibfield
  {journal} {\bibinfo  {journal} {Physical Review Letters}\ }\textbf {\bibinfo
  {volume} {123}},\ \bibinfo {pages} {253901} (\bibinfo {year}
  {2019})}\BibitemShut {NoStop}%
\bibitem [{\citenamefont {Kupriianov}\ \emph {et~al.}(2019)\citenamefont
  {Kupriianov}, \citenamefont {Xu}, \citenamefont {Sayanskiy}, \citenamefont
  {Dmitriev}, \citenamefont {Kivshar},\ and\ \citenamefont
  {Tuz}}]{kupriianov2019metasurface}%
  \BibitemOpen
  \bibfield  {author} {\bibinfo {author} {\bibfnamefont {A.~S.}\ \bibnamefont
  {Kupriianov}}, \bibinfo {author} {\bibfnamefont {Y.}~\bibnamefont {Xu}},
  \bibinfo {author} {\bibfnamefont {A.}~\bibnamefont {Sayanskiy}}, \bibinfo
  {author} {\bibfnamefont {V.}~\bibnamefont {Dmitriev}}, \bibinfo {author}
  {\bibfnamefont {Y.~S.}\ \bibnamefont {Kivshar}}, \ and\ \bibinfo {author}
  {\bibfnamefont {V.~R.}\ \bibnamefont {Tuz}},\ }\href@noop {} {\bibfield
  {journal} {\bibinfo  {journal} {Physical Review Appl.}\ }\textbf {\bibinfo
  {volume} {12}},\ \bibinfo {pages} {014024} (\bibinfo {year}
  {2019})}\BibitemShut {NoStop}%
\bibitem [{\citenamefont {Mikheeva}\ \emph {et~al.}(2019)\citenamefont
  {Mikheeva}, \citenamefont {Koshelev}, \citenamefont {Choi}, \citenamefont
  {Kruk}, \citenamefont {Lumeau}, \citenamefont {Abdeddaim}, \citenamefont
  {Voznyuk}, \citenamefont {Enoch},\ and\ \citenamefont
  {Kivshar}}]{mikheeva2019photosensitive}%
  \BibitemOpen
  \bibfield  {author} {\bibinfo {author} {\bibfnamefont {E.}~\bibnamefont
  {Mikheeva}}, \bibinfo {author} {\bibfnamefont {K.}~\bibnamefont {Koshelev}},
  \bibinfo {author} {\bibfnamefont {D.-Y.}\ \bibnamefont {Choi}}, \bibinfo
  {author} {\bibfnamefont {S.}~\bibnamefont {Kruk}}, \bibinfo {author}
  {\bibfnamefont {J.}~\bibnamefont {Lumeau}}, \bibinfo {author} {\bibfnamefont
  {R.}~\bibnamefont {Abdeddaim}}, \bibinfo {author} {\bibfnamefont
  {I.}~\bibnamefont {Voznyuk}}, \bibinfo {author} {\bibfnamefont
  {S.}~\bibnamefont {Enoch}}, \ and\ \bibinfo {author} {\bibfnamefont
  {Y.}~\bibnamefont {Kivshar}},\ }\href@noop {} {\bibfield  {journal} {\bibinfo
   {journal} {Optics Express}\ }\textbf {\bibinfo {volume} {27}},\ \bibinfo
  {pages} {33847} (\bibinfo {year} {2019})}\BibitemShut {NoStop}%
\bibitem [{\citenamefont {Koshelev}\ \emph
  {et~al.}(2019{\natexlab{b}})\citenamefont {Koshelev}, \citenamefont {Tang},
  \citenamefont {Li}, \citenamefont {Choi}, \citenamefont {Li},\ and\
  \citenamefont {Kivshar}}]{koshelev2019nonlinear}%
  \BibitemOpen
  \bibfield  {author} {\bibinfo {author} {\bibfnamefont {K.}~\bibnamefont
  {Koshelev}}, \bibinfo {author} {\bibfnamefont {Y.}~\bibnamefont {Tang}},
  \bibinfo {author} {\bibfnamefont {K.}~\bibnamefont {Li}}, \bibinfo {author}
  {\bibfnamefont {D.-Y.}\ \bibnamefont {Choi}}, \bibinfo {author}
  {\bibfnamefont {G.}~\bibnamefont {Li}}, \ and\ \bibinfo {author}
  {\bibfnamefont {Y.}~\bibnamefont {Kivshar}},\ }\href@noop {} {\bibfield
  {journal} {\bibinfo  {journal} {ACS Photonics}\ } (\bibinfo {year}
  {2019}{\natexlab{b}})}\BibitemShut {NoStop}%
\bibitem [{\citenamefont {Xu}\ \emph {et~al.}(2019)\citenamefont {Xu},
  \citenamefont {Zangeneh~Kamali}, \citenamefont {Huang}, \citenamefont
  {Rahmani}, \citenamefont {Smirnov}, \citenamefont {Camacho-Morales},
  \citenamefont {Ma}, \citenamefont {Zhang}, \citenamefont {Woolley},
  \citenamefont {Neshev} \emph {et~al.}}]{xu2019dynamic}%
  \BibitemOpen
  \bibfield  {author} {\bibinfo {author} {\bibfnamefont {L.}~\bibnamefont
  {Xu}}, \bibinfo {author} {\bibfnamefont {K.}~\bibnamefont {Zangeneh~Kamali}},
  \bibinfo {author} {\bibfnamefont {L.}~\bibnamefont {Huang}}, \bibinfo
  {author} {\bibfnamefont {M.}~\bibnamefont {Rahmani}}, \bibinfo {author}
  {\bibfnamefont {A.}~\bibnamefont {Smirnov}}, \bibinfo {author} {\bibfnamefont
  {R.}~\bibnamefont {Camacho-Morales}}, \bibinfo {author} {\bibfnamefont
  {Y.}~\bibnamefont {Ma}}, \bibinfo {author} {\bibfnamefont {G.}~\bibnamefont
  {Zhang}}, \bibinfo {author} {\bibfnamefont {M.}~\bibnamefont {Woolley}},
  \bibinfo {author} {\bibfnamefont {D.}~\bibnamefont {Neshev}},  \emph
  {et~al.},\ }\href@noop {} {\bibfield  {journal} {\bibinfo  {journal}
  {Advanced Science}\ ,\ \bibinfo {pages} {1802119}} (\bibinfo {year}
  {2019})}\BibitemShut {NoStop}%
\bibitem [{\citenamefont {He}\ \emph {et~al.}(2018)\citenamefont {He},
  \citenamefont {Guo}, \citenamefont {Feng}, \citenamefont {Xu},\ and\
  \citenamefont {Miroshnichenko}}]{he2018toroidal}%
  \BibitemOpen
  \bibfield  {author} {\bibinfo {author} {\bibfnamefont {Y.}~\bibnamefont
  {He}}, \bibinfo {author} {\bibfnamefont {G.}~\bibnamefont {Guo}}, \bibinfo
  {author} {\bibfnamefont {T.}~\bibnamefont {Feng}}, \bibinfo {author}
  {\bibfnamefont {Y.}~\bibnamefont {Xu}}, \ and\ \bibinfo {author}
  {\bibfnamefont {A.~E.}\ \bibnamefont {Miroshnichenko}},\ }\href@noop {}
  {\bibfield  {journal} {\bibinfo  {journal} {Phys. Rev. B}\ }\textbf {\bibinfo
  {volume} {98}},\ \bibinfo {pages} {161112} (\bibinfo {year}
  {2018})}\BibitemShut {NoStop}%
\bibitem [{\citenamefont {Shen}\ \emph {et~al.}(2019)\citenamefont {Shen},
  \citenamefont {Zhou}, \citenamefont {Ge}, \citenamefont {Hu},\ and\
  \citenamefont {Lu}}]{shen2019liquid}%
  \BibitemOpen
  \bibfield  {author} {\bibinfo {author} {\bibfnamefont {Z.-X.}\ \bibnamefont
  {Shen}}, \bibinfo {author} {\bibfnamefont {S.-H.}\ \bibnamefont {Zhou}},
  \bibinfo {author} {\bibfnamefont {S.-J.}\ \bibnamefont {Ge}}, \bibinfo
  {author} {\bibfnamefont {W.}~\bibnamefont {Hu}}, \ and\ \bibinfo {author}
  {\bibfnamefont {Y.-Q.}\ \bibnamefont {Lu}},\ }\href@noop {} {\bibfield
  {journal} {\bibinfo  {journal} {Applied Physics Letters}\ }\textbf {\bibinfo
  {volume} {114}},\ \bibinfo {pages} {041106} (\bibinfo {year}
  {2019})}\BibitemShut {NoStop}%
\bibitem [{\citenamefont {Parry}\ \emph {et~al.}(2017)\citenamefont {Parry},
  \citenamefont {Komar}, \citenamefont {Hopkins}, \citenamefont {Campione},
  \citenamefont {Liu}, \citenamefont {Miroshnichenko}, \citenamefont {Nogan},
  \citenamefont {Sinclair}, \citenamefont {Brener},\ and\ \citenamefont
  {Neshev}}]{parry2017active}%
  \BibitemOpen
  \bibfield  {author} {\bibinfo {author} {\bibfnamefont {M.}~\bibnamefont
  {Parry}}, \bibinfo {author} {\bibfnamefont {A.}~\bibnamefont {Komar}},
  \bibinfo {author} {\bibfnamefont {B.}~\bibnamefont {Hopkins}}, \bibinfo
  {author} {\bibfnamefont {S.}~\bibnamefont {Campione}}, \bibinfo {author}
  {\bibfnamefont {S.}~\bibnamefont {Liu}}, \bibinfo {author} {\bibfnamefont
  {A.~E.}\ \bibnamefont {Miroshnichenko}}, \bibinfo {author} {\bibfnamefont
  {J.}~\bibnamefont {Nogan}}, \bibinfo {author} {\bibfnamefont {M.~B.}\
  \bibnamefont {Sinclair}}, \bibinfo {author} {\bibfnamefont {I.}~\bibnamefont
  {Brener}}, \ and\ \bibinfo {author} {\bibfnamefont {D.~N.}\ \bibnamefont
  {Neshev}},\ }\href@noop {} {\bibfield  {journal} {\bibinfo  {journal}
  {Applied Physics Letters}\ }\textbf {\bibinfo {volume} {111}},\ \bibinfo
  {pages} {053102} (\bibinfo {year} {2017})}\BibitemShut {NoStop}%
\bibitem [{\citenamefont {Zhou}\ \emph {et~al.}(2018)\citenamefont {Zhou},
  \citenamefont {Liu}, \citenamefont {Ban}, \citenamefont {Li}, \citenamefont
  {Huang}, \citenamefont {Xia}, \citenamefont {Wang},\ and\ \citenamefont
  {Zhan}}]{zhou2018tunable}%
  \BibitemOpen
  \bibfield  {author} {\bibinfo {author} {\bibfnamefont {C.}~\bibnamefont
  {Zhou}}, \bibinfo {author} {\bibfnamefont {G.}~\bibnamefont {Liu}}, \bibinfo
  {author} {\bibfnamefont {G.}~\bibnamefont {Ban}}, \bibinfo {author}
  {\bibfnamefont {S.}~\bibnamefont {Li}}, \bibinfo {author} {\bibfnamefont
  {Q.}~\bibnamefont {Huang}}, \bibinfo {author} {\bibfnamefont
  {J.}~\bibnamefont {Xia}}, \bibinfo {author} {\bibfnamefont {Y.}~\bibnamefont
  {Wang}}, \ and\ \bibinfo {author} {\bibfnamefont {M.}~\bibnamefont {Zhan}},\
  }\href@noop {} {\bibfield  {journal} {\bibinfo  {journal} {Applied Physics
  Letters}\ }\textbf {\bibinfo {volume} {112}},\ \bibinfo {pages} {101904}
  (\bibinfo {year} {2018})}\BibitemShut {NoStop}%
\bibitem [{\citenamefont {Li}\ \emph {et~al.}()\citenamefont {Li},
  \citenamefont {Gupta}, \citenamefont {Zhang}, \citenamefont {Wang},
  \citenamefont {Chen}, \citenamefont {Singh}, \citenamefont {Han},\ and\
  \citenamefont {Zhang}}]{liactive}%
  \BibitemOpen
  \bibfield  {author} {\bibinfo {author} {\bibfnamefont {Q.}~\bibnamefont
  {Li}}, \bibinfo {author} {\bibfnamefont {M.}~\bibnamefont {Gupta}}, \bibinfo
  {author} {\bibfnamefont {X.}~\bibnamefont {Zhang}}, \bibinfo {author}
  {\bibfnamefont {S.}~\bibnamefont {Wang}}, \bibinfo {author} {\bibfnamefont
  {T.}~\bibnamefont {Chen}}, \bibinfo {author} {\bibfnamefont {R.}~\bibnamefont
  {Singh}}, \bibinfo {author} {\bibfnamefont {J.}~\bibnamefont {Han}}, \ and\
  \bibinfo {author} {\bibfnamefont {W.}~\bibnamefont {Zhang}},\ }\href@noop {}
  {\bibinfo  {journal} {Advanced Materials Technologies}\ ,\ \bibinfo {pages}
  {1900840}}\BibitemShut {NoStop}%
\bibitem [{\citenamefont {Cao}\ \emph {et~al.}(2016)\citenamefont {Cao},
  \citenamefont {Bao}, \citenamefont {Mao}, \citenamefont {Zhang},
  \citenamefont {Novitsky}, \citenamefont {Nieto-Vesperinas},\ and\
  \citenamefont {Qiu}}]{cao2016controlling}%
  \BibitemOpen
\bibfield  {journal} {  }\bibfield  {author} {\bibinfo {author} {\bibfnamefont
  {T.}~\bibnamefont {Cao}}, \bibinfo {author} {\bibfnamefont {J.}~\bibnamefont
  {Bao}}, \bibinfo {author} {\bibfnamefont {L.}~\bibnamefont {Mao}}, \bibinfo
  {author} {\bibfnamefont {T.}~\bibnamefont {Zhang}}, \bibinfo {author}
  {\bibfnamefont {A.}~\bibnamefont {Novitsky}}, \bibinfo {author}
  {\bibfnamefont {M.}~\bibnamefont {Nieto-Vesperinas}}, \ and\ \bibinfo
  {author} {\bibfnamefont {C.-W.}\ \bibnamefont {Qiu}},\ }\href@noop {}
  {\bibfield  {journal} {\bibinfo  {journal} {Acs Photonics}\ }\textbf
  {\bibinfo {volume} {3}},\ \bibinfo {pages} {1934} (\bibinfo {year}
  {2016})}\BibitemShut {NoStop}%
\bibitem [{\citenamefont {Chu}\ \emph {et~al.}(2016)\citenamefont {Chu},
  \citenamefont {Tseng}, \citenamefont {Chen}, \citenamefont {Wu},
  \citenamefont {Chen}, \citenamefont {Wang}, \citenamefont {Chen},
  \citenamefont {Hsieh}, \citenamefont {Wu}, \citenamefont {Sun} \emph
  {et~al.}}]{chu2016active}%
  \BibitemOpen
  \bibfield  {author} {\bibinfo {author} {\bibfnamefont {C.~H.}\ \bibnamefont
  {Chu}}, \bibinfo {author} {\bibfnamefont {M.~L.}\ \bibnamefont {Tseng}},
  \bibinfo {author} {\bibfnamefont {J.}~\bibnamefont {Chen}}, \bibinfo {author}
  {\bibfnamefont {P.~C.}\ \bibnamefont {Wu}}, \bibinfo {author} {\bibfnamefont
  {Y.-H.}\ \bibnamefont {Chen}}, \bibinfo {author} {\bibfnamefont {H.-C.}\
  \bibnamefont {Wang}}, \bibinfo {author} {\bibfnamefont {T.-Y.}\ \bibnamefont
  {Chen}}, \bibinfo {author} {\bibfnamefont {W.~T.}\ \bibnamefont {Hsieh}},
  \bibinfo {author} {\bibfnamefont {H.~J.}\ \bibnamefont {Wu}}, \bibinfo
  {author} {\bibfnamefont {G.}~\bibnamefont {Sun}},  \emph {et~al.},\
  }\href@noop {} {\bibfield  {journal} {\bibinfo  {journal} {Laser \& Photonics
  Reviews}\ }\textbf {\bibinfo {volume} {10}},\ \bibinfo {pages} {986}
  (\bibinfo {year} {2016})}\BibitemShut {NoStop}%
\bibitem [{\citenamefont {Wang}\ \emph {et~al.}(2016)\citenamefont {Wang},
  \citenamefont {Rogers}, \citenamefont {Gholipour}, \citenamefont {Wang},
  \citenamefont {Yuan}, \citenamefont {Teng},\ and\ \citenamefont
  {Zheludev}}]{wang2016optically}%
  \BibitemOpen
  \bibfield  {author} {\bibinfo {author} {\bibfnamefont {Q.}~\bibnamefont
  {Wang}}, \bibinfo {author} {\bibfnamefont {E.~T.}\ \bibnamefont {Rogers}},
  \bibinfo {author} {\bibfnamefont {B.}~\bibnamefont {Gholipour}}, \bibinfo
  {author} {\bibfnamefont {C.-M.}\ \bibnamefont {Wang}}, \bibinfo {author}
  {\bibfnamefont {G.}~\bibnamefont {Yuan}}, \bibinfo {author} {\bibfnamefont
  {J.}~\bibnamefont {Teng}}, \ and\ \bibinfo {author} {\bibfnamefont {N.~I.}\
  \bibnamefont {Zheludev}},\ }\href@noop {} {\bibfield  {journal} {\bibinfo
  {journal} {Nature Photonics}\ }\textbf {\bibinfo {volume} {10}},\ \bibinfo
  {pages} {60} (\bibinfo {year} {2016})}\BibitemShut {NoStop}%
\bibitem [{\citenamefont {Li}\ \emph {et~al.}(2019{\natexlab{b}})\citenamefont
  {Li}, \citenamefont {Zhou}, \citenamefont {Ban}, \citenamefont {Wang},
  \citenamefont {Lu},\ and\ \citenamefont {Wang}}]{li2019active}%
  \BibitemOpen
  \bibfield  {author} {\bibinfo {author} {\bibfnamefont {S.}~\bibnamefont
  {Li}}, \bibinfo {author} {\bibfnamefont {C.}~\bibnamefont {Zhou}}, \bibinfo
  {author} {\bibfnamefont {G.}~\bibnamefont {Ban}}, \bibinfo {author}
  {\bibfnamefont {H.}~\bibnamefont {Wang}}, \bibinfo {author} {\bibfnamefont
  {H.}~\bibnamefont {Lu}}, \ and\ \bibinfo {author} {\bibfnamefont
  {Y.}~\bibnamefont {Wang}},\ }\href@noop {} {\bibfield  {journal} {\bibinfo
  {journal} {Journal of Physics D: Applied Physics}\ }\textbf {\bibinfo
  {volume} {52}},\ \bibinfo {pages} {095106} (\bibinfo {year}
  {2019}{\natexlab{b}})}\BibitemShut {NoStop}%
\bibitem [{\citenamefont {Tian}\ \emph {et~al.}(2019)\citenamefont {Tian},
  \citenamefont {Luo}, \citenamefont {Yang}, \citenamefont {Ding},
  \citenamefont {Qu}, \citenamefont {Zhao}, \citenamefont {Qiu},\ and\
  \citenamefont {Bozhevolnyi}}]{tian2019active}%
  \BibitemOpen
  \bibfield  {author} {\bibinfo {author} {\bibfnamefont {J.}~\bibnamefont
  {Tian}}, \bibinfo {author} {\bibfnamefont {H.}~\bibnamefont {Luo}}, \bibinfo
  {author} {\bibfnamefont {Y.}~\bibnamefont {Yang}}, \bibinfo {author}
  {\bibfnamefont {F.}~\bibnamefont {Ding}}, \bibinfo {author} {\bibfnamefont
  {Y.}~\bibnamefont {Qu}}, \bibinfo {author} {\bibfnamefont {D.}~\bibnamefont
  {Zhao}}, \bibinfo {author} {\bibfnamefont {M.}~\bibnamefont {Qiu}}, \ and\
  \bibinfo {author} {\bibfnamefont {S.~I.}\ \bibnamefont {Bozhevolnyi}},\
  }\href@noop {} {\bibfield  {journal} {\bibinfo  {journal} {Nature Commun.}\
  }\textbf {\bibinfo {volume} {10}},\ \bibinfo {pages} {396} (\bibinfo {year}
  {2019})}\BibitemShut {NoStop}%
\bibitem [{\citenamefont {Qu}\ \emph {et~al.}(2017)\citenamefont {Qu},
  \citenamefont {Li}, \citenamefont {Du}, \citenamefont {Cai}, \citenamefont
  {Lu},\ and\ \citenamefont {Qiu}}]{qu2017dynamic}%
  \BibitemOpen
  \bibfield  {author} {\bibinfo {author} {\bibfnamefont {Y.}~\bibnamefont
  {Qu}}, \bibinfo {author} {\bibfnamefont {Q.}~\bibnamefont {Li}}, \bibinfo
  {author} {\bibfnamefont {K.}~\bibnamefont {Du}}, \bibinfo {author}
  {\bibfnamefont {L.}~\bibnamefont {Cai}}, \bibinfo {author} {\bibfnamefont
  {J.}~\bibnamefont {Lu}}, \ and\ \bibinfo {author} {\bibfnamefont
  {M.}~\bibnamefont {Qiu}},\ }\href@noop {} {\bibfield  {journal} {\bibinfo
  {journal} {Laser \& Photonics Reviews}\ }\textbf {\bibinfo {volume} {11}},\
  \bibinfo {pages} {1700091} (\bibinfo {year} {2017})}\BibitemShut {NoStop}%
\bibitem [{\citenamefont {Karvounis}\ \emph {et~al.}(2016)\citenamefont
  {Karvounis}, \citenamefont {Gholipour}, \citenamefont {MacDonald},\ and\
  \citenamefont {Zheludev}}]{karvounis2016all}%
  \BibitemOpen
  \bibfield  {author} {\bibinfo {author} {\bibfnamefont {A.}~\bibnamefont
  {Karvounis}}, \bibinfo {author} {\bibfnamefont {B.}~\bibnamefont
  {Gholipour}}, \bibinfo {author} {\bibfnamefont {K.~F.}\ \bibnamefont
  {MacDonald}}, \ and\ \bibinfo {author} {\bibfnamefont {N.~I.}\ \bibnamefont
  {Zheludev}},\ }\href@noop {} {\bibfield  {journal} {\bibinfo  {journal}
  {Applied Physics Letters}\ }\textbf {\bibinfo {volume} {109}},\ \bibinfo
  {pages} {051103} (\bibinfo {year} {2016})}\BibitemShut {NoStop}%
\bibitem [{\citenamefont {Tseng}\ \emph {et~al.}(2012)\citenamefont {Tseng},
  \citenamefont {Wu}, \citenamefont {Sun}, \citenamefont {Chang}, \citenamefont
  {Chen}, \citenamefont {Chu}, \citenamefont {Chen}, \citenamefont {Zhou},
  \citenamefont {Huang}, \citenamefont {Yen} \emph
  {et~al.}}]{tseng2012fabrication}%
  \BibitemOpen
  \bibfield  {author} {\bibinfo {author} {\bibfnamefont {M.~L.}\ \bibnamefont
  {Tseng}}, \bibinfo {author} {\bibfnamefont {P.~C.}\ \bibnamefont {Wu}},
  \bibinfo {author} {\bibfnamefont {S.}~\bibnamefont {Sun}}, \bibinfo {author}
  {\bibfnamefont {C.~M.}\ \bibnamefont {Chang}}, \bibinfo {author}
  {\bibfnamefont {W.~T.}\ \bibnamefont {Chen}}, \bibinfo {author}
  {\bibfnamefont {C.~H.}\ \bibnamefont {Chu}}, \bibinfo {author} {\bibfnamefont
  {P.-L.}\ \bibnamefont {Chen}}, \bibinfo {author} {\bibfnamefont
  {L.}~\bibnamefont {Zhou}}, \bibinfo {author} {\bibfnamefont {D.-W.}\
  \bibnamefont {Huang}}, \bibinfo {author} {\bibfnamefont {T.-J.}\ \bibnamefont
  {Yen}},  \emph {et~al.},\ }\href@noop {} {\bibfield  {journal} {\bibinfo
  {journal} {Laser \& Photonics Reviews}\ }\textbf {\bibinfo {volume} {6}},\
  \bibinfo {pages} {702} (\bibinfo {year} {2012})}\BibitemShut {NoStop}%
\bibitem [{\citenamefont {Pandian}\ \emph {et~al.}(2007)\citenamefont
  {Pandian}, \citenamefont {Kooi}, \citenamefont {Palasantzas}, \citenamefont
  {De~Hosson},\ and\ \citenamefont {Pauza}}]{pandian2007nanoscale}%
  \BibitemOpen
  \bibfield  {author} {\bibinfo {author} {\bibfnamefont {R.}~\bibnamefont
  {Pandian}}, \bibinfo {author} {\bibfnamefont {B.~J.}\ \bibnamefont {Kooi}},
  \bibinfo {author} {\bibfnamefont {G.}~\bibnamefont {Palasantzas}}, \bibinfo
  {author} {\bibfnamefont {J.~T.}\ \bibnamefont {De~Hosson}}, \ and\ \bibinfo
  {author} {\bibfnamefont {A.}~\bibnamefont {Pauza}},\ }\href@noop {}
  {\bibfield  {journal} {\bibinfo  {journal} {Advanced Materials}\ }\textbf
  {\bibinfo {volume} {19}},\ \bibinfo {pages} {4431} (\bibinfo {year}
  {2007})}\BibitemShut {NoStop}%
\bibitem [{\citenamefont {Palik}(1998)}]{palik1998handbook}%
  \BibitemOpen
  \bibfield  {author} {\bibinfo {author} {\bibfnamefont {E.~D.}\ \bibnamefont
  {Palik}},\ }\href@noop {} {\emph {\bibinfo {title} {Handbook of optical
  constants of solids}}},\ Vol.~\bibinfo {volume} {3}\ (\bibinfo  {publisher}
  {Academic press},\ \bibinfo {year} {1998})\BibitemShut {NoStop}%
\bibitem [{\citenamefont {Fedotov}\ \emph {et~al.}(2010)\citenamefont
  {Fedotov}, \citenamefont {Papasimakis}, \citenamefont {Plum}, \citenamefont
  {Bitzer}, \citenamefont {Walther}, \citenamefont {Kuo}, \citenamefont
  {Tsai},\ and\ \citenamefont {Zheludev}}]{fedotov2010spectral}%
  \BibitemOpen
  \bibfield  {author} {\bibinfo {author} {\bibfnamefont {V.~A.}\ \bibnamefont
  {Fedotov}}, \bibinfo {author} {\bibfnamefont {N.}~\bibnamefont
  {Papasimakis}}, \bibinfo {author} {\bibfnamefont {E.}~\bibnamefont {Plum}},
  \bibinfo {author} {\bibfnamefont {A.}~\bibnamefont {Bitzer}}, \bibinfo
  {author} {\bibfnamefont {M.}~\bibnamefont {Walther}}, \bibinfo {author}
  {\bibfnamefont {P.}~\bibnamefont {Kuo}}, \bibinfo {author} {\bibfnamefont
  {D.}~\bibnamefont {Tsai}}, \ and\ \bibinfo {author} {\bibfnamefont
  {N.}~\bibnamefont {Zheludev}},\ }\href@noop {} {\bibfield  {journal}
  {\bibinfo  {journal} {Physical review letters}\ }\textbf {\bibinfo {volume}
  {104}},\ \bibinfo {pages} {223901} (\bibinfo {year} {2010})}\BibitemShut
  {NoStop}%
\bibitem [{\citenamefont {Yang}\ \emph {et~al.}(2014)\citenamefont {Yang},
  \citenamefont {Kravchenko}, \citenamefont {Briggs},\ and\ \citenamefont
  {Valentine}}]{yang2014all}%
  \BibitemOpen
  \bibfield  {author} {\bibinfo {author} {\bibfnamefont {Y.}~\bibnamefont
  {Yang}}, \bibinfo {author} {\bibfnamefont {I.~I.}\ \bibnamefont
  {Kravchenko}}, \bibinfo {author} {\bibfnamefont {D.~P.}\ \bibnamefont
  {Briggs}}, \ and\ \bibinfo {author} {\bibfnamefont {J.}~\bibnamefont
  {Valentine}},\ }\href@noop {} {\bibfield  {journal} {\bibinfo  {journal}
  {Nature communications}\ }\textbf {\bibinfo {volume} {5}},\ \bibinfo {pages}
  {5753} (\bibinfo {year} {2014})}\BibitemShut {NoStop}%
\bibitem [{\citenamefont {Campione}\ \emph {et~al.}(2016)\citenamefont
  {Campione}, \citenamefont {Liu}, \citenamefont {Basilio}, \citenamefont
  {Warne}, \citenamefont {Langston}, \citenamefont {Luk}, \citenamefont
  {Wendt}, \citenamefont {Reno}, \citenamefont {Keeler}, \citenamefont {Brener}
  \emph {et~al.}}]{campione2016broken}%
  \BibitemOpen
  \bibfield  {author} {\bibinfo {author} {\bibfnamefont {S.}~\bibnamefont
  {Campione}}, \bibinfo {author} {\bibfnamefont {S.}~\bibnamefont {Liu}},
  \bibinfo {author} {\bibfnamefont {L.~I.}\ \bibnamefont {Basilio}}, \bibinfo
  {author} {\bibfnamefont {L.~K.}\ \bibnamefont {Warne}}, \bibinfo {author}
  {\bibfnamefont {W.~L.}\ \bibnamefont {Langston}}, \bibinfo {author}
  {\bibfnamefont {T.~S.}\ \bibnamefont {Luk}}, \bibinfo {author} {\bibfnamefont
  {J.~R.}\ \bibnamefont {Wendt}}, \bibinfo {author} {\bibfnamefont {J.~L.}\
  \bibnamefont {Reno}}, \bibinfo {author} {\bibfnamefont {G.~A.}\ \bibnamefont
  {Keeler}}, \bibinfo {author} {\bibfnamefont {I.}~\bibnamefont {Brener}},
  \emph {et~al.},\ }\href@noop {} {\bibfield  {journal} {\bibinfo  {journal}
  {Acs Photonics}\ }\textbf {\bibinfo {volume} {3}},\ \bibinfo {pages} {2362}
  (\bibinfo {year} {2016})}\BibitemShut {NoStop}%
\end{thebibliography}%

\end{document}